\input harvmac
\noblackbox
\input epsf

\newcount\figno
\figno=0
\def\fig#1#2#3{
\par\begingroup\parindent=0pt\leftskip=1cm\rightskip=1cm\parindent=0pt
\baselineskip=11pt

\global\advance\figno by 1
\midinsert
\epsfxsize=#3
\centerline{\epsfbox{#2}}
\vskip 12pt
\centerline{{\bf Figure \the\figno:} #1}\par
\endinsert\endgroup\par}
\def\figlabel#1{\xdef#1{\the\figno}}

\def\risunok#1#2#3{\vskip 15pt
\global\advance\figno by 1
\centerline{\epsfbox{#1}}
\vskip 10pt
\centerline{{\bf Fig. #3: } #2}
\vskip 15pt }
\def\kartinka#1#2#3{
\par\begingroup\parindent=0pt\leftskip=1cm\rightskip=1cm\parindent=0pt
\baselineskip=11pt

\global\advance\figno by 1
\midinsert
\centerline{\epsfbox{#1}}
\vskip 12pt
\centerline{{\bf Fig. #3: }#2}\par
\endinsert\endgroup\par}

\def\np#1#2#3{Nucl. Phys. {\bf B#1} (#2) #3}
\def\pl#1#2#3{Phys. Lett. {\bf B#1} (#2) #3}
\def\prl#1#2#3{Phys. Rev. Lett.{\bf #1} (#2) #3}
\def\physrev#1#2#3{Phys. Rev. {\bf D#1} (#2) #3}
\def\ap#1#2#3{Ann. Phys. {\bf #1} (#2) #3}

\def\cmp#1#2#3{Comm. Math. Phys. {\bf #1} (#2) #3}


\font\cmss=cmss10
\font\cmsss=cmss10 at 7pt
\def\rlx{\relax\leavevmode}
\def\inbar{\vrule height1.5ex width.4pt depth0pt}
\def\IC{\relax\,\hbox{$\inbar\kern-.3em{\rm C}$}}
\def\IN{\relax{\rm I\kern-.18em N}}
\def\IP{\relax{\rm I\kern-.18em P}}
\def\ZZ{\rlx\leavevmode\ifmmode\mathchoice{\hbox{\cmss Z\kern-.4em Z}}
 {\hbox{\cmss Z\kern-.4em Z}}{\lower.9pt\hbox{\cmsss Z\kern-.36em Z}}
 {\lower1.2pt\hbox{\cmsss Z\kern-.36em Z}}\else{\cmss Z\kern-.4em
 Z}\fi}
\def\IZ{\relax\ifmmode\mathchoice
{\hbox{\cmss Z\kern-.4em Z}}{\hbox{\cmss Z\kern-.4em Z}}
{\lower.9pt\hbox{\cmsss Z\kern-.4em Z}}
{\lower1.2pt\hbox{\cmsss Z\kern-.4em Z}}\else{\cmss Z\kern-.4em
Z}\fi}

\def\narrowplus{\kern -.04truein + \kern -.03truein}
\def\narrowminus{- \kern -.04truein}
\def\narrowminussub{\kern -.02truein - \kern -.01truein}

\def\kh{K\"{a}hler }

\def\b{{\beta}}
\def\a{{\alpha}}

\def\e{{\epsilon}}

\def\r{{\rightarrow}}

\def\frac#1#2{{#1\over #2}}

\def\RV{{R_\Vert}}

\def\IZ{\relax\ifmmode\mathchoice
{\hbox{\cmss Z\kern-.4em Z}}{\hbox{\cmss Z\kern-.4em Z}}
{\lower.9pt\hbox{\cmsss Z\kern-.4em Z}}
{\lower1.2pt\hbox{\cmsss Z\kern-.4em Z}}\else{\cmss Z\kern-.4em
Z}\fi}
\def\IB{\relax{\rm I\kern-.18em B}}
\def\IC{{\relax\hbox{$\inbar\kern-.3em{\rm C}$}}}
\def\ID{\relax{\rm I\kern-.18em D}}
\def\IE{\relax{\rm I\kern-.18em E}}
\def\IF{\relax{\rm I\kern-.18em F}}
\def\IG{\relax\hbox{$\inbar\kern-.3em{\rm G}$}}
\def\IGa{\relax\hbox{${\rm I}\kern-.18em\Gamma$}}
\def\IH{\relax{\rm I\kern-.18em H}}
\def\II{\relax{\rm I\kern-.18em I}}
\def\IK{\relax{\rm I\kern-.18em K}}
\def\IP{\relax{\rm I\kern-.18em P}}

\font\cmss=cmss10 \font\cmsss=cmss10 at 7pt
\def\IR{\relax{\rm I\kern-.18em R}}

\def\S{{\Sigma}}

\def\D{{\cal D}}

\def\G{{ \rm \bf g}^*}
\def\gg{{ \cal G  }}
\def\ra{\rightarrow}
\def\s{ \sigma}
\def\wT{{\widehat T}}
\def\wS{{\widehat S}}
\def\oz{{\overline{z}}}

%

%
%
\def\eqnn#1{\xdef #1{(\secsym\the\meqno)}\writedef{#1\leftbracket#1}%
\global\advance\meqno by1\wrlabeL#1}
\def\eqna#1{\xdef #1##1{\hbox{$(\secsym\the\meqno##1)$}}
\writedef{#1\numbersign1\leftbracket#1{\numbersign1}}%
\global\advance\meqno by1\wrlabeL{#1$\{\}$}}
\def\eqn#1#2{\xdef #1{(\secsym\the\meqno)}\writedef{#1\leftbracket#1}%
\global\advance\meqno by1$$#2\eqno#1\eqlabeL#1$$}



\lref\rimpurity{S. Sethi, hep-th/9710005.}
\lref\rnew{O. Ganor and S. Sethi, hep-th/9712071, JHEP 01 (1998) 007.}
\lref\rwati{W. Taylor IV, hep-th/9611042, \pl{394}{1997}{283}; hep-th/9801182.}
\lref\rori{O. Ganor, S. Ramgoolam, and W. Taylor IV, hep-th/9611202,
\np{492}{1997}{191}.}
\lref\rgreg{A. Gerasimov, G. Moore, and S. Shatashvili, unpublished.}
\lref\rsim{D. Tsimpis, to appear.}
\lref\rhitchin{N.J. Hitchin, \cmp{83}{1982}{579}.}
\lref\rHKLR{N.J. Hitchin, A. Karlhede, U. Lindstr\"om, and M. Ro\v{c}ek,
\cmp{108}{1987}{535}.}
\lref\rmdf{M. Douglas, hep-th/9512077; hep-th/9604198.}
\lref\rwitten{E. Witten, hep-th/9511030, \np{460}{1996}{541}.}
\lref\rnahm{W. Nahm, ``Self-Dual Monopoles and Calorons,'' {\it Lect. Notes in
Phys. 201,}
(Springer, New York, 1984).}
\lref\rhurt{J. Hurtubise and M.K. Murray, \cmp{122}{1989}{35}.}
\lref\rdiaconescu{D.-E. Diaconescu, hep-th/9608163, \np{503}{1997}{220}. }
\lref\rgm{H. Garland and M.K. Murray, \cmp{120}{1988}{335}.}
\lref\rcg{E. Corrigan and P. Goddard, \ap{154}{1984}{253}.}
\lref\rleeyi{K. Lee and P. Yi, hep-th/9702107, \physrev{56}{1997}{3711}.}
\lref\rdancer{A.S. Dancer, \cmp{158}{1993}{545}.}
\lref\rwest{P. West, {\it Introduction to Supersymmetry and Supergravity,}
(World Scientific Publishing, Singapore, 1990). }
\lref\rpetr{P. Horava and E. Witten, hep-th/9603142, \np{475}{1996}{94}.}
\lref\rweinberg{K. Lee, E. Weinberg, and P. Yi, hep-th/9602167,
\physrev{54}{1996}{1636}.}
\lref\rkraan{K. Lee, hep-th/9802012; K. Lee and C. Lu, hep-th/9802108\semi
 T. C. Kraan and P. van Baal, hep-th/9802049.}
\lref\rlimit{N. Seiberg hep-th/9710009, \prl{79}{1997}{3577}\semi
A. Sen, hep-th/9709220.}
\lref\rahn{C. Ahn and B.-H. Lee, hep-th/9803069.}
\lref\rfour{E. Witten, hep-th/9703166, \np{500}{1997}{3}. }
\lref\rdonagi{R. Donagi and E. Witten,  hep-th/9510101,
\np{460}{1996}{299}.}
\lref\rberk{M. Berkooz, hep-th/9802069; O. Aharony, M. Berkooz,
and N. Seiberg, hep-th/9712117.}
\lref\rnon{A. Connes, M. Douglas, and A. Schwarz, hep-th/9711162\semi
M. Douglas and C. Hull, hep-th/9711165.}
\lref\rnik{N. Nekrasov and A. Schwarz, hep-th/9802068.}
\lref\rnahms{W. Nahm, ``The Construction of All Self-Dual Monopoles
by the ADHM Method,'' {\it Monopoles in Quantum Field Theory}, eds.
N. S. Craigie et al., (World
Scientific, Singapore, 1982). }
\lref\rnikt{A. Gorsky and N. Nekrasov, hep-th/9401021.}
\lref\rmor{H. Itoyama and A. Morozov, hep-th/9511126, \np{477}{1996}{855}.}
\lref\rkap{A. Kapustin, to appear.}


\lref\rpol{J. Polchinski, ``TASI Lectures on D-Branes,''
hep-th/9611050\semi J. Polchinski, S. Chaudhuri and C. Johnson,
``Notes on D-Branes,'' hep-th/9602052. }
\lref\rBFSS{T. Banks, W. Fischler, S. H. Shenker, and L. Susskind,
hep-th/9610043, Phys. Rev. {\bf D55} (1997) 5112.}
\lref\rwtensor{E. Witten,
hep-th/9507121.}
\lref\rstensor{A. Strominger, ``Open P-Branes,'' hep-th/9512059,
\pl{383}{1996}{44}.}
\lref\rsdecoupled{N. Seiberg, ``New Theories in Six Dimensions and
Matrix Description of M-theory on $T^5$ and $T^5/\IZ_2$,''
hep-th/9705221.}
\lref\rashoke{ A. Sen, ``Kaluza-Klein Dyons in String
Theory,'' hep-th/9705212; ``A Note on Enhanced Gauge Symmetries in
M and String Theory,'' hep-th/9707123; ``Dynamics of Multiple
Kaluza-Klein Monopoles in M and String Theory,'' hep-th/9707042.}
\lref\kutetal{D. Berenstein, R. Corrado and J. Distler, ``On the
Moduli Spaces of M(atrix)-Theory Compactifications,''
hep-th/9704087\semi  S. Elitzur, A. Giveon, D. Kutasov and
E. Rabinovici, ``Algebraic Aspects of Matrix Theory on $T^d$,''
hep-th/9707217.}
\lref\rtwoform{J. P. Gauntlett and D. Lowe, ``Dyons and S-Duality in
N=4 Supersymmetric Gauge Theory,'' hep-th/9601085,
\np{472}{1996}{194}\semi K. Lee, E. Weinberg and P. Yi,
``Electromagnetic Duality and $SU(3)$ Monopoles,'' hep-th/9601097,
\pl{376}{1996}{97}.}
\lref\rmoore{A. Losev, G. Moore, and S. Shatashvili, ``M \& m's ,''
hep-th/9707250.}
\lref\rbrunner{I. Brunner and A. Karch, ``Matrix Description of
M-theory on $T^6$,'' hep-th/9707259.}
\lref\rDVV{R. Dijkgraaf, E. Verlinde and H. Verlinde, ``BPS Spectrum
of the Five-Brane and Black Hole Entropy,'' hep-th/9603126,
\np{486}{1997}{77}; ``BPS Quantization of the Five-Brane,''
hep-th/9604055, \np{486}{1997}{89}.}
\lref\rsixbrane{P. Townsend, ``The Eleven Dimensional Supermembrane
Revisited,'' hep-th/9501068, \pl{350}{1995}{184}.}
\lref\rmultitn{S. Hawking, ``Gravitational Instantons,''
Phys. Lett. {\bf 60A} (1977) 81\semi
G. Gibbons and S. Hawking, ``Classification of Gravitational Instanton
Symmetries,'' Comm. Math. Phys. {\bf 66} (1979) 291\semi
R. Sorkin, ``Kaluza-Klein Monopole,''
\prl{51}{1983}{87}\semi D. Gross and M. Perry, ``Magnetic Monopoles in
Kaluza-Klein Theories,'' \np{226}{1983}{29}.}
\lref\rmIIB{P. Aspinwall, ``Some Relationships Between Dualities in
String Theory,'' hep-th/9508154, Nucl. Phys. Proc. Suppl. {\bf 46}
(1996) 30\semi J. Schwarz, ``The Power of M Theory,''
hep-th/9510086, \pl{367}{1996}{97}. }
\lref\rgeneralsixbrane{ C. Hull,  ``Gravitational Duality, Branes and
Charges,'' hep-th/9705162\semi E. Bergshoeff, B. Janssen, and
T. Ortin, ``Kaluza-Klein Monopoles and Gauged Sigma Models,''
hep-th/9706117\semi Y. Imamura, ``Born-Infeld Action and Chern-Simons
Term {}from Kaluza-Klein Monopole in M-theory,'' hep-th/9706144.}
\lref\rbd{M. Berkooz and M. Douglas, hep-th/9610236, \pl{395}{1997}{196}.}
\lref\rbraneswith{M. Douglas, ``Branes within Branes,''
hep-th/9512077.}
\lref\rquantumfive{O. Aharony, M. Berkooz, S. Kachru, N. Seiberg, and
E. Silverstein, hep-th/9707079, A.T.M.P. {\bf 1} (1998) 148.}
\lref\rstringfive{E. Witten, hep-th/9707093, JHEP 07 (1997) 003.}
\lref\rSS{S. Sethi and L. Susskind, ``Rotational Invariance in the
M(atrix) Formulation of Type IIB Theory,'' hep-th/9702101,
\pl{400}{1997}{265}.}
\lref\rBS{T. Banks and N. Seiberg, ``Strings from Matrices,''
hep-th/9702187, \np{497}{1997}{41}.}
\lref\rgilad{A. Hanany and G. Lifschytz, ``M(atrix) Theory on $T^6$
and a m(atrix) Theory Description of KK Monopoles,'' hep-th/9708037.}
\lref\rreview{N. Seiberg, ``Notes on Theories with 16 Supercharges,''
hep-th/9705117.}
\lref\rDVVstring{R. Dijkgraaf, E. Verlinde and H. Verlinde, ``Matrix
String Theory,'' hep-th/9703030.}
\lref\rprobes{M. Douglas, ``Gauge Fields and D-branes,''
hep-th/9604198.}
\lref\doumoo{M. Douglas and G. Moore,  hep-th/9603167.}
\lref\rfischler{M. Douglas, ``Enhanced Gauge Symmetry in M(atrix)
Theory,'' hep-th/9612126\semi
W. Fischler and A. Rajaraman, ``M(atrix) String Theory
on K3,'' hep-th/9704123\semi
C. Johnson and R. Myers, ``Aspects of Type IIB Theory on ALE Spaces,''
hep-th/9610140, \physrev{55}{1997}{6382}\semi
D.-E. Diaconescu and J. Gomis, ``Duality in Matrix Theory
and Three Dimensional Mirror Symmetry,'' hep-th/9707019.}
\lref\rsprobes{N. Seiberg, ``Gauge Dynamics And Compactification To
Three Dimensions,'' hep-th/9607163, \pl{384}{1996}{81}}
\lref\rswthree{N. Seiberg and E. Witten, ``Gauge Dynamics and
Compactifications to Three Dimensions,'' hep-th/9607163.}
\lref\rthroat{D.-E. Diaconescu and N. Seiberg, ``The Coulomb Branch of
$(4,4)$ Supersymmetric Field Theories in Two Dimensions,''
hep-th/9707158. }
\lref\rTduality{T. Banks, M. Dine, H. Dykstra and W. Fischler,
``Magnetic Monopole Solutions of String Theory,''
\pl{212}{1988}{45}\semi C. Hull and P. Townsend, ``Unity of
Superstring Dualities,'' hep-th/9410167, \np{438}{109}{1995}\semi
H. Ooguri, C. Vafa, ``Two Dimensional Black Hole and Singularities of
Calabi-Yau Manifolds,'' Nucl.Phys. {\bf B463} (1996) 55,
hep-th/9511164\semi D. Kutasov, ``Orbifolds and Solitons,'' Phys. Lett
{\bf B383} (1996) 48, hep-th/9512145\semi H. Ooguri and C. Vafa,
``Geometry of N=1 Dualities in Four Dimensions,'' hep-th/9702180.}
\lref\raps{P. Argyres, R. Plesser and N. Seiberg, ``The Moduli Space
of N=2 SUSY QCD and Duality in N=1 SUSY QCD,'' hep-th/9603042,
\np{471}{1996}{159}.}
\lref\rgms{O. Ganor, D. Morrison and N. Seiberg, ``Branes, Calabi-Yau
Spaces, and Toroidal Compactification of the N=1 Six Dimensional $E_8$
Theory,'' hep-th/9610251, \np{487}{1997}{93}.}
\lref\rchs{ C.G. Callan, J.A. Harvey, A. Strominger, ``Supersymmetric
String Solitons,'' hep-th/9112030, \np{359}{1991}{611}\semi S.-J. Rey,
in ``The  Proc. of the Tuscaloosa Workshop
1989,'' 291; Phys. Rev. {\bf D43} (1991) 526; S.-J. Rey, In DPF Conf.
1991, 876.}
\lref\rmotl{L. Motl, ``Proposals on nonperturbative superstring
interactions,'' hep-th/9701025.}

\Title{\vbox{\hbox{hep-th/9804027}\hbox{IASSNS-HEP-98/33}}}
{\vbox{\centerline{The Higgs Branch of Impurity Theories}}}
\smallskip
\centerline{Anton Kapustin\footnote{$^1$} {kapustin@sns.ias.edu}
 and Savdeep Sethi\footnote{$^2$} {sethi@sns.ias.edu}  }
\vskip 0.12in
\medskip\centerline{\it School of Natural Sciences}
\centerline{\it Institute for Advanced Study}\centerline{\it
Princeton, NJ
08540, USA}

\vskip 1in

We consider supersymmetric gauge theories with impurities in various
dimensions. These
systems arise in the study of intersecting branes. Unlike conventional
gauge theories, the Higgs branch of an impurity theory can have compact
directions. For models with eight supercharges, the Higgs branch is a
hyper\kh manifold given by the moduli
space of solutions of certain differential equations. These equations are the
dimensional
reductions of self-duality equations with boundary conditions determined by the
impurities. They can also be interpreted as Nahm transforms of self-duality
equations on toroidally compactified spaces. We discuss the application of 
our results to the light-cone formulation of Yang-Mills theories and to the
solution of certain N=2 d=4 gauge theories.

\vskip 0.1in
\Date{4/98}

\newsec{Introduction}

Intersecting brane configurations are important in both string theory and
matrix theory \rBFSS. When the branes involved are Dirichlet, we should be
able to describe the dynamics of the intersecting brane configuration in terms 
of
a gauge theory with some degrees of freedom localized at the intersection.
These
degrees of freedom come from strings stretching between the intersecting
branes.
We
will call theories with localized degrees of freedom `impurity theories'
\refs{
\rimpurity, \rnew}.

The aim of this paper is to analyze  the impurity theories that
arise from the
intersection of partially compactified D-branes. For example, the theory
describing
the dynamics of $N$
D0-branes in the presence of $k$ D4-branes wrapped on a circle should have two
branches. On the Higgs branch, the fundamental hypermultiplets coming from the
D0-D4 strings have expectation
values. The D0-branes then cannot leave the D4-branes since the moduli
parametrizing
motion away from the D4-branes are massive. The D0-branes essentially become
instantons
living in the D4-branes. The Higgs branch of this impurity theory
must have compact directions since it should be describing the moduli space of
instantons on $\IR^3 \times S^1$. On the Coulomb branch, we can give
expectation
values to the moduli parametrizing
motion away from the D4-branes. On this branch, the fundamental hypermultiplets
are
massive. Similarly, when the D4-branes are wrapped on a more general
space,
the Higgs branch should describe the instantons on this space and
therefore can have compact directions.

Our goal is to show how
compact moduli spaces appear in impurity theories. The models we consider have
eight
supercharges and non-chiral impurities. In these cases, the Higgs branch is
hyper\kh and not renormalized by quantum corrections. In the following section,
we derive the Lagrangian for particular impurity theories and analyze the Higgs
branch. In section three, we explain how the Higgs branch is related to the 
Nahm transform of the self-duality equations on $\IR^3\times S^1$ and 
$\IR^2\times T^2$. We conclude by explaining how our results are applicable to 
matrix formulations of Yang-Mills theories, and to the solution of certain N=2
d=4 theories. 

It would be interesting
to
extend this analysis to impurity theories with fewer supersymmetries and to 
theories  with
localized chiral matter. Such systems arise, for example, in the study of type
I
D1-branes in the
presence of D5-branes \rmdf\ wrapped on tori. Related results have
been obtained by \refs{\rgreg, \rsim}.

\newsec{Supersymmetric Vacua of Impurity Systems}
\subsec{The supersymmetric Lagrangian}

The primary example that we will study is the system of $N$ D0-branes probing
$k$ D4-branes wrapped on tori. To derive the impurity Lagrangian, we use the
approach pioneered in \refs{\rwati, \rori}, and extended to the
impurity case in \refs{\rimpurity, \rnew}. Let us start with the case of
four-branes on $\IR^4$. The theory on the D0-branes is then a $U(N)$ quantum
mechanics with
$k$
fundamental hypermultiplets and one adjoint hypermultiplet. For simplicity, we
will take
D4-branes to be coincident. Separating the four-branes amounts
to turning on bare masses for the fundamental hypermultiplets.

The field content can be obtained by dimensionally reducing an
$N=2$ gauge theory in $d=4$. Our conventions are similar to those in \rwest.
The global symmetry group is $Spin(5)\times SU(2)_R\times SU(2)$,
and a vector multiplet contains five scalars which we call $Y^i,i=1,\ldots, 5$,
and a pair of symplectic Majorana fermions $\lambda^\a$, $\a=1,2$ transforming
as a
doublet under the $SU(2)_R$ symmetry (we are using a `four-dimensional'
language
to describe the fermions). The adjoint hypermultiplet
has two complex scalars $ H^\a,$ $\a=1,2$ forming a doublet under $SU(2)_R$ and
a doublet with respect to the other $SU(2)$. It also contains
a Dirac fermion $\psi$.
Lastly, the $k$ fundamental hypermultiplets each contain
two complex scalars $Q^{\a p}$, where $\a=1,2$ are the $SU(2)_R$ indices
and $p=1,\ldots,k$ are the flavor indices, together with associated fermions.
The adjoint hypermultiplet encodes degrees of freedom from the $0-0$ strings.
Its four real scalars parametrize the motion of the D0-branes along the
world-volume of
the D4-branes. We will write all real adjoint fields as anti-Hermitian
matrices. The bosonic part of the Lagrangian
is a sum of three terms,
$$ L = L_1 + L_2 + L_3.$$
The first contains the vector multiplet kinetic terms,
\eqn\vector{L_1 = \int{ dt  \left({1\over 2} |D_0 Y^i|^2 -{1\over 2}
\sum_{i<j}
|[Y^i, Y^j]|^2  \right). }}
The covariant derivative is the usual one,
$$ D_0 = \partial_t - A_0. $$
The second piece contains the kinetic terms of the adjoint hypermultiplet,
\eqn\adjoint{L_2 = \int{ dt  \left(  |D_0 H^\a |^2 - \sum
| [Y^i, H^\a] |^2   \right),  }}
The final term contains the fundamental hypermultiplets
and $\D$-terms,
\eqn\fundamental{L_3 = \int{ dt  \left( |D_0 Q^{\a}|^2
- \sum | Y^i Q^{\a}|^2 + {1\over 2} |\D|^2  + \Tr\ i\D_\b^\a\,
\left( \, [ H_\a, H^{\dagger \b}] +
 Q_\a\otimes Q^{\dagger \b}  \, \right) \right),  }}
where flavor indices have been suppressed. In eq. \fundamental,
$$ \D_\b^\a = \D^a (\s^a)_\b^\a, $$
where $\D^a$ $a=1,2,3$ is a triplet of auxiliary fields in the adjoint. The
$SU(2)_R$ indices
are moved up and down using $\epsilon_{\a\b}$ so that $\D^{\a\b}$ is symmetric.
The tensor product refers to the $U(N)$ (color) indices.

Let us pick $ X^1= {\sqrt 2}\, {\rm Re}\, H_1$ as the longitudinal direction
which
we wish to
compactify. We then take the system and all its translates along $X^1$. This
describes an
array of an infinite number of D0-branes.
The gauge group is now infinite-dimensional and we quotient by the symmetry
group generated by
translations along $X^1$ by $2 \pi i R_1$ \refs{\rwati, \rori}. Recall that
$X^1$ is anti-Hermitian. More explicitly,
we
impose the constraints:
\eqn\constraints{ \eqalign{ & Y^i_{nm} = Y^i_{(n-1)(m-1)}, \cr
& H'_{nm} = H'_{(n-1)(m-1)}, \cr
& \D_{nm} = \D_{(n-1)(m-1)}, \cr
& X^1_{nm} = X^1_{(n-1)(m-1)} \qquad n \neq m, \cr
& X^1_{nn} = X^1_{(n-1)(n-1)} + 2\pi i\, R_1, \cr}}
where $H'$ excludes $X^1$ and the subscripts $n,m$ label the translate. Note
that each component, say $Y_{nm}$, is still an $N \times N$ matrix transforming
in the adjoint representation of our original $U(N)$ gauge group.
As usual, it proves convenient to perform a Fourier transform on all the
adjoint
fields, e.g.
\eqn\Fourier{ Y^i(x^1)=\sum_n Y^i_{n0} e^{2\pi inR_1x_1}.}
This promotes $Y, H',\D$ to fields living on a circle $\wS^1$ of radius
${\widehat R}_1 = 1/{2 \pi R_1}$. As for $X^1$, its Fourier transform is a
differential
operator rather than a function:
\eqn\conn{ X^1(x^1)=\, \partial_1 - A_1(x_1).}
The fundamental hypermultiplets coming from the D0-D4 strings are treated
differently
since the D4-branes are longitudinal. The $p^{\rm th}$ hypermultiplets $Q^p$
obeys,
\eqn\hyperconstr{Q_n^p =e^{-2\pi iR_1ns_p}Q_0^p,}
where $s_1,\ldots,s_k$ are real numbers. They parametrize the Wilson lines
for the $U(k)$ gauge theory on the D4-branes. The fundamental hypermultiplets
are therefore not promoted to fields on $\wS^1$. We will see in a moment that
the
$p^{\rm th}$ hypermultiplet is in fact `localized' at $x^1=s_p$.

After substituting the expressions for $Y$ and $H$
in terms of their Fourier transforms  into eq. \vector\ and eq. \adjoint,
we obtain:
\eqn\L{\eqalign {& L_1+L_2 = R_1 \int{ dt dx_1  \left[{1\over 2}|F_{01}|^2+
{1\over 2} |D_0 Y^i|^2 -
{1\over 2} |D_1 Y^i|^2- {1\over 2} \sum_{i<j} |[Y^i, Y^j]|^2  \right.} \cr
& +\left.  |D_0 {\rm Im}\ H_1|^2+|D_0 H_2|^2 - \sum_i \left(
| [Y^i, H_2 ] |^2  + | [Y^i, {\rm Im}\ H_1]|^2\right)\right] . \cr }}
Here $F_{01}=-[D_0,D_1].$
The more interesting terms in the Lagrangian involve the couplings to $Q$. The
terms in $L_3$ become,
\eqn\kinimpurity{ \eqalign{ & L_3 = R_1\int{ dx_1 dt \,\left[ \sum_{p=1}^k
\left(
{1\over R_1}\delta(x_1-s_p)\left( |D_0 Q^{\a p}|^2 - \sum | Y^i(x_1) Q^{\a
p}|^2
\right)\right) \right.} \cr
& \left.+{1\over 2} |\D|^2 + \Tr\ i\D_\b^\a\left([H_\a,H^{\dagger \b}]+{1\over
R_1}
\sum_{p=1}^k\delta(x_1-s_p) Q_\a^p\otimes Q^{\dagger p\b}\right)\right], \cr }}
where it is understood that ${\rm Re}\ H_1$ is replaced by ${1\over
\sqrt{2}}(\partial_1
-A_1)$, and the $\delta$-function on $\wS^1$ is defined so that
$\oint dx^1 \delta(x_1)=1$.
After integrating out the auxiliary field $\D$, the resulting Lagrangian will
contain terms proportional to $\delta(0)$  which may seem problematic.
However,
this should not be very surprising in theories with boundary interactions. For
example,
M theory on $S^1/\IZ_2$ which is relevant to the strong-coupling limit of the
$E_8\times E_8$ heterotic string, has similar divergences appearing in the
Lagrangian
\rpetr.
As we shall see, the  $\delta$-function terms in $\D$-terms are crucial if we
are to obtain
a Higgs branch with the properties that we expect on physical grounds.

\subsec{ $\IR^3 \times S^1$}

Let us first consider the case of $N$ D0-branes in the presence of
$k$ D4-branes wrapped on a circle of radius $R_1$. On the Higgs branch,
we set $Y=0$ but allow the $Q$ hypermultiplets to have
non-zero
expectation values. With this choice, the supersymmetric variation of
the `squarks', the superpartners of $Q$, automatically vanishes. To ensure
that we have a supersymmetric vacuum, we must
force the supersymmetric variation of the gluinos to vanish as well. This
reduces to the
condition that the $\D$-terms vanish.
Note, however, that this condition is now a differential equation because ${\rm
Re}\ H_1$
is
a differential operator.
The vacuum is therefore not spatially homogeneous.
To write the $\D$-flatness condition in a more familiar form, we denote:
 \eqn\definitions{ \eqalign{& T_0 = - A_1, \quad T_1 =-\sqrt{2}{\rm Im} H_1,
   \quad T_2+iT_3 = -\sqrt{2} H_2 , \cr &  Q=Q_1, \quad \tilde{Q} =
(Q_2)^\dagger.}}
We can then rewrite the equations to get,
\eqn\Dflatness{\eqalign{ &
{d T_1\over dx_1}+[T_0,T_1]+[T_2,T_3]=-{i\over R_1}\sum_{p=1}^k \delta(x_1-s_p)
\left(Q^p\otimes Q^{\dagger p}-\tilde{Q}^{\dagger
p}\otimes\tilde{Q}^p\right),\cr
& {d T_2\over dx_1}+[T_0,T_2]+[T_3,T_1]=-{i\over R_1}\sum_{p=1}^k
\delta(x_1-s_p)
\left(-i Q^p\otimes\tilde{Q}^p+i\tilde{Q}^{\dagger p}\otimes Q^{\dagger
p}\right),\cr
& {d T_3\over dx_1}+[T_0,T_3]+[T_1,T_2]=-{i\over R_1}\sum_{p=1}^k
\delta(x_1-s_p)
\left(Q^p\otimes\tilde{Q}^p+\tilde{Q}^{\dagger p}\otimes Q^{\dagger
p}\right).\cr}}
These equations without the right-hand side source terms are known as
Nahm equations. They
play an important role in the construction of self-dual monopoles and
calorons.\foot{Otherwise known as instantons on $\IR^3\times S^1$.}
We will discuss these equations in more detail in section three.

If the VEVs of the $Q$ and $\tilde{Q}$ hypermultiplets are non-zero,
eqs. \Dflatness\ force
$T_i,i=1,2,3,$
to have step-like discontinuities at the points $s_1,\ldots,s_k$. Let
us see how this picture is interpreted in terms of the D-branes.
After T-duality along the $S^1$, we are studying a system of $N$ D1-branes
wrapped on $\wS^1$ with $k$ D3-branes located at $k$ points $s_1,\ldots,s_k$ on
$\wS^1$. The VEVs of the $T_i$ fields parametrize the positions of the
D1-branes in the directions $x^2,x^3,x^4$
parallel to the D3-branes. The discontinuity in $T_i$ at $x_1=s_p$ is
now
interpreted as D1-strings breaking at the $p^{\rm th}$ D3-brane, as pictured
in figure 2.1. This breaking prevents the D1-strings from moving off in the
directions
perpendicular to the D3-branes. This means that fields in the vector multiplet
are massive.

\kartinka{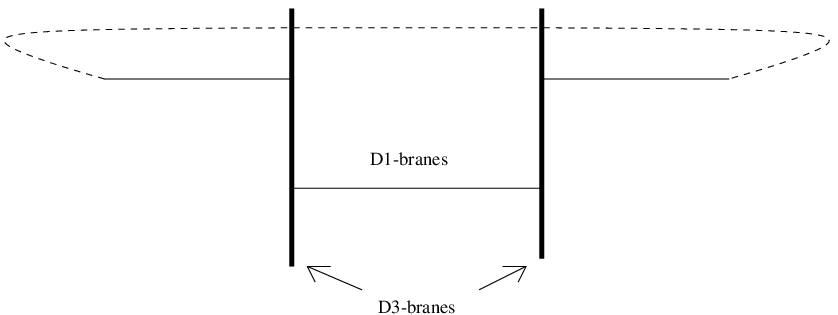}{Jumps in $T_i$}{2.1}

Naturally, we should consider the solutions of the $\D$-flatness equations
modulo
$U(N)$ gauge transformations. The moduli space of solutions to the $\D$-term
equations should
then coincide with the moduli space of $N$ $U(k)$ instantons on $ \IR^3 \times
S^1$
with the fixed Wilson lines. We will give another explanation of this in
section
three.

\subsec{ $\IR^2 \times T^2$}

Let us compactify another direction along the D4-branes, say
$X^2 = {\sqrt 2}\, {\rm Im} H_1$, on a circle of radius $R_2$.
The procedure described before goes through in the
same way. The adjoint degrees of freedom become fields on the dual torus
$\wT^2$; in particular,
\eqn\conntwo{ X^2 \r \,  \partial_2 -  A_2.}
The fundamental degrees of freedom are localized at points of $\wT^2$
$z_1,\ldots,z_k$
which encode the $U(k)$ Wilson lines on $T^2$.
Let us define $A_\oz={1\over 2}(A_1+iA_2), \Phi={1\over \sqrt{2}} H^{\dagger
2}$. Then
the Higgs branch is described by the moduli space of solutions of the
equations,
\eqn\hitchin{ \eqalign{ & F_{z\oz}-[\Phi,\Phi^\dagger]={1\over
2R_1R_2}\sum_{p=1}^k
\delta^2(z-z_p)\left( Q^p\otimes Q^{\dagger p} -\tilde{Q}^{\dagger p}
\otimes\tilde{Q}^p\right),\cr
& \overline{D}\Phi=-{1\over 2R_1R_2}\sum_{p=1}^k \delta^2(z-z_p)
Q^p\otimes\tilde{Q}^p. \cr}}
Here $F_{z\oz}=\partial A_\oz-{\overline{\partial}} A_z-[A_z,A_\oz]$ and $
\overline{D}=\overline{\partial}-A_\oz.$

These equations are Hitchin equations on the dual torus
$\wT^2$ with impurities localized at points of $\wT^2$. Recalling that these
equations
describe $N$ D0-branes stuck to $k$ D4-branes wrapped around $\IR^2\times T^2$,
we expect that
the moduli space of solutions of eq. \hitchin\ coincides with the moduli
space of $N$ $U(k)$ instantons on $ \IR^2 \times T^2$ with prescribed Wilson
lines.

If we compactify more than two directions, we run into a problem. The
$\D$-flatness
conditions will then involve $ \delta^3(x)$ or $\delta^4(x)$ terms for $T^3$ or
$T^4$
compactifications, respectively. These seem to be too singular source terms for
the
equations to admit any solutions with non-zero $Q$ expectation values. There is
a
physical way of seeing why
these cases might be problematic. Consider the case of D4-branes wrapped around
$T^3$.
After T-dualizing D4-branes on $T^3$, we obtain impurity D1-branes
while our `probe' D0-branes become D3-branes wrapped on $\wT^3$.
We want to be able to localize the
position of the impurities at points on  $ {\widehat T}^3$. This means that we
want to give expectation values to scalars in the $1+1$-dimensional theory on
the D1-branes. However, the theory on D1-branes does not have a moduli space,
so
we can
no longer regard the position of the D1-branes as fixed. Our attempt to do so
leads
to equations for the Higgs branch with no nontrivial solutions. The same
problem
occurs
for $T^4$ compactifications. To deal with these cases, we should be doing
quantum
mechanics on the impurity world-volume.

\subsec{Fayet-Iliopoulos parameters}
Since the gauge group of our impurity theories is $U(N)$, it is possible to
modify
the $\D$-flatness conditions by Fayet-Iliopoulos (FI) terms. More precisely,
the gauge group is the group of maps from $\wS^1$ or $\wT^2$ to $U(N)$, so FI
terms can be $functions$ on $\wS^1$ or $\wT^2$.
In the presence of FI terms the $\D$-flatness conditions
are modified to
$$ \D_i=\xi_i,$$
where both $\D$ and $\xi$ are functions on $\wS^1$ or $\wT^2$.
Naively, this seems to introduce a continuum of deformation parameters.
However, it is easy to see that in fact only the average of $\xi$ over
$\wS^1$ or $\wT^2$ matters. For example, in the case considered in section
2.2,
a change of variables $T_i(x^1)\r T_i(x^1)+h_i(x^1)$ shifts $\xi_i$ by
$d h_i/dx^1$. Since such a change of variables leaves the moduli space
unchanged, we see that the only `gauge-invariant' information is contained
in $\oint dx^1\xi_i$. Similarly, in the case considered in section 2.3
the only invariant is $\int d^2z\, \xi_i(z,\oz)$.

Recall that
in the case of instantons on $\IR^4$ the modification of  the ADHM equations by
FI
terms corresponds to a $B$-field flux \rberk. This FI deformation has an
interpretation in terms of non-commutative geometry \rnon\ as instantons
on a noncommutative generalization of $\IR^4$ \rnik.
As explained in the next section, Nahm
and Hitchin equations with impurities are nothing but the Nahm transform
of instantons on $\IR^3\times S^1$ and $\IR^2\times T^2$, respectively.
The FI terms should again correspond to a background B-field
flux. These deformed moduli spaces should therefore be interpreted as
describing instantons on a non-commutative $\IR^3\times S^1$ and
$\IR^2\times T^2$. That there is a single three-component FI parameter
corresponds to the fact that there are three closed self-dual 2-forms
on $\IR^3\times S^1$ and $\IR^2\times T^2$.

\subsec{More general gauge groups}

There is no reason why the preceeding analysis cannot be
extended to more general groups. Most straightforward are the two cases that
naturally arise in string theory with orientifolds. Instead of starting with
a system of just D4-branes and D0-branes, we can place either an $O4^{-}$ or
$O4^{+}$ orientifold plane parallel to the D4-branes without breaking more
supersymmetry.

In the former case, we
have either an even or an odd number of D4-branes. The D4-branes have an
$SO(k)$ gauge symmetry while the D0-branes have an $Sp(N)$ gauge group.
There are $k$ half-hypermultiplets in the fundamental and a hypermultiplet in
the antisymmetric representation. In the $O4^{+}$ case, we can only have
an even number of D4-branes giving an $Sp(k)$ gauge symmetry. In this case,
the D0-branes have an $SO(N)$ gauge symmetry with $k/2$ hypermultiplets in the
vector representation and a hypermultiplet in the symmetric representation.

After compactifying a longitudinal direction, we will end up with a theory
of $N$ D-strings with two orientifold 3-planes and $k$ D3-branes.
Away from the orientifold points, the gauge group on the D-strings is still
 $U(N)$. The only effect of the orientifold projection is to relate the
matrices
$T_0(x_1),T_i(x_1)$, as well as the impurity degrees of freedom,
at points on the circle related by $x^1\r -x^1$. See \refs{\rhurt,\rahn}\
for the analogous discussion in the case of monopoles. Similarly,
in the case of compactification on $T^2$ we obtain the same equations
(Hitchin equations) on $\wT^2$, with an additional identification under
$z\r -z$. It would be interesting to investigate the corresponding moduli
spaces
in more detail for both $S^1$ and $T^2$.

\newsec{The Reduction of Self-Duality Equations}

\subsec{Hyper\kh quotients}

In a conventional gauge theory with gauge group $G$, the Higgs branch is
determined by
solving the algebraic equation of $\D$-flatness,
$$ \D = 0. $$
The moduli space of solutions has no compact directions for these models. It
has
a
natural hyper\kh metric because the equation $\D=0$ defines a
hyper\kh moment map for the group of gauge transformations $G$. Let us recall
how this
comes about. A hyper\kh manifold is a Riemannian manifold with metric $g$ and
three
complex structures $I,J$ and $K$ satisfying three conditions:

\vskip 0.1in
\noindent
1. $IJ = -IJ = K$ so $I, J, K$ form an algebra of quaternions.

\noindent
2. $I, J$ and $K$ are covariantly constant with respect to $g$.

\noindent
3. $g$ is Hermitian with respect to all of the complex structures.
\vskip 0.1in

{}From these conditions, it follows that there are three \kh forms associated
to
each complex
structure, $ \omega_1, \omega_2$ and $\omega_3$. Suppose $X$ is a hyper\kh
manifold admitting
an action of the Lie group $G$ preserving the metric and the three complex
structures. For the
case of gauge theories, $X$ is the space of hypermultiplet VEVs.
The group $G$
preserves the three \kh forms, and any one of the \kh forms $\omega_i$ defines
a
symplectic
structure on $X$. Therefore the action of $G$ is generated by a Hamiltonian
$\mu_i$
valued in $ \G$, the dual of the Lie algebra of $G$. Together these three
Hamiltonians define a moment map $\mu$ from $X$ to $\G \otimes \IR^3$. It
follows from
these definitions that the level set of $\mu$, $\mu^{-1}(0)$, is invariant with
respect to $G$. Then according to the theorem
of \rHKLR, the quotient $ \mu^{-1} (0)/G $ is a hyper\kh space.

Let us consider a particular example of hyper\kh quotient. Let the group $U(N)$
act by
left multiplication on the flat hyper\kh manifold $\IH^N$. This action
commutes
with
right multiplication by $I, J, K$ and therefore preserves all three complex
structures.
The corresponding moment map is most compactly written if we think of $\IH^N$
as
$\, \IC^N\times \,\IC^N$ with coordinates $(Q,\tilde{Q})$. Then the moment map
is
given by
\eqn\unmap{\mu_1=i(Q\otimes Q^\dagger
-\tilde{Q}^\dagger\otimes\tilde{Q}),\qquad
\mu_2-i\mu_3=2i\ Q\otimes \tilde{Q}.}
In this particular case, the hyper\kh quotient $\mu^{-1}(0)/G$ is a single
point
$Q=\tilde{Q}=0.$

In the case of impurity theories, we have seen that the Higgs branch is
described
by the moduli
spaces of solutions of differential rather than algebraic equations. In the
cases
that we considered, these equations
are Nahm  or Hitchin equations with
point-like impurities. As expected on general grounds,
these moduli spaces are hyper\kh spaces. A way to see this is to note that
these
equations can be regarded as moment map equations for an infinite-dimensional
group of
gauge transformations $ \gg$ \rhitchin.
For the case considered in section 2.2, where space was a circle, we
obtained
Nahm equations.
The corresponding infinite-dimensional gauge group $\gg$ is the loop group of
$G$, i.e. it
is the group of maps from $\wS^1$ to $G$.
This group acts on an
infinite-dimensional flat hyper\kh manifold consisting of all quadruplets
$(T_0,T_1,T_2,T_3),$ where
$T_0$ is a $U(N)$ connection on $\wS^1$ and $T_i$ are adjoint-valued fields on
$\wS^1$.
The model in section 2.3 lived on $\wT^2$, and its Higgs branch was
described
by
Hitchin equations. In this case, $\gg$ is the group of maps from $\wT^2$ to
$G$.
It acts on the space of quadruplets $(A_z,A_\oz,\Phi,\Phi^\dagger),$
where $A_z dz+A_\oz d\oz$ is a $U(N)$ connection on $\wT^2$ and $\Phi dz
+\Phi^\dagger d\oz$ is
an adjoint-valued 1-form on $ T^2$. In both cases the theorem of \rHKLR\
guarantees that the
space of solutions modulo gauge transformations is a hyper\kh space.

To account for the impurity degrees of freedom, we
have to enlarge the initial configuration space by including the fundamental
hypermultiplets
living at $k$ points $z_1,\ldots,z_k$ on either $\wS^1$ or $\wT^2$. Each
fundamental
hypermultiplet
takes values in $\IH^N$, and we have just seen that there is an action of
$G=U(N)$
on $\IH^N$ preserving the hyper\kh structure. Thus there is a natural action of
$\gg$ on the enlarged
space of variables, with $g(z)\in\gg$ acting on the $p^{\rm th}$ hypermultiplet
$Q_p$
by
$Q_p\ra g(z_p) Q_p$. The moment map then contains an extra contribution
proportional
to a sum of delta functions. Let us denote the moment map for the bulk
variables
by $\mu_\gg,$
and the moment map for the degrees of freedom localized at $z_p$ by
$\rho_p.$
The explicit
form of $\rho_p$ for the action of $U(N)$ on $\IH^N$ is given in eq. \unmap.
Then the
combined moment map $\mu_T$ is given by,
\eqn\momentmap{\mu_T = \mu_\gg + \sum_p \delta (z-z_p)\rho_p.}
The equations \Dflatness\ and \hitchin\ for the supersymmetric vacua of our
impurity models are exactly of this form.

\subsec{The relation to solutions of self-duality equations}

The impurity model discussed in section 2.2 ($N$ D0-branes bound to $k$
parallel D4-branes compactified on a circle) describes $N$ $U(k)$
instantons on $\IR^3\times S^1$. Instantons on $\IR^3\times S^1$ are also known
as calorons \rnahm.
Thus we infer that the moduli space of calorons coincides with the
moduli space of Nahm equations on a circle with point-like impurities.
In this case, we have obtained a known result which can be derived in several
different
ways \refs{\rnahm,\rgm}. The novelty in our derivation is that we obtain the
answer
by studying the dynamics of an impurity system. This kind of approach was
first used in \rwitten\ where type I 5-branes in flat space were
studied.
In that case, the gauge theory was an ordinary gauge theory without impurities,
and so
the equation $\D=0$ was algebraic. The $\D$-flatness conditions were identical
to
the equations arising in the ADHM construction of instantons on $\IR^4$. This
agreed with
the expectation that the Higgs branch of type I 5-branes corresponds to
finite size instantons in the transverse $\IR^4$.

In the case considered in section 2.3 (D0-branes bound to D4-branes
compactified on $T^2$) we obtained Hitchin equations rather than
Nahm equations. We are going to argue that the moduli space of Hitchin
equations on the dual torus $\wT^2$ with extra degrees of freedom in the
fundamental,
localized at punctures, correctly describes the moduli space of $U(k)$
instantons on
$\IR^2\times T^2$. It seems that instantons on $\IR^2\times T^2$ have not been
discussed in the literature, so in this section we intend to fill this gap.
But before doing so, we recall how Nahm equations come about in the case of
calorons.

The correspondence between solutions of self-duality equations on $\IR^3\times
S^1$
and Nahm equations on $\wS^1$ is known as the Nahm transform 
\refs{\rnahm, \rnahms}. The Nahm
transform
maps
every solution of the former into a solution of the latter in the following
way.
Consider a self-dual $U(k)$ connection $A$ on $\IR^3\times S^1$ with an
asymptotic
Wilson line at spatial infinity breaking $U(k)$ down to $U(1)^k$. The
topological
invariants characterizing such a connection are its second Chern class $N$ and
its $k-1$ magnetic charges with respect to the unbroken $U(1)$ factors;
the charge in
the diagonal $U(1)$ subgroup is always zero. We consider
normalizable solutions of the Dirac equation,
$$\sigma\cdot D_A\psi(x,s)=is\psi(x,s),$$
where $\psi$ is in the fundamental representation of $U(k)$ and $s$ is a real
parameter. Because $e^{is}$ can be regarded as an auxiliary $U(1)$ Wilson
line around $S^1$,  $s$ is a periodic variable living on the dual circle
$\wS^1$.
The number of normalizable
zero modes $\psi(x,s)$ is locally constant in $s$, but may have jumps at points
$e^{i s}=W_p$ where $W_p,\ p=1,\ldots,k,$ are the eigenvalues of the $U(k)$
Wilson line at
spatial infinity. We next define four $s$-dependent matrices:
$$ \eqalign{ T_i(s) & =i\int \psi^\dagger x_i \psi\ d^4x,\ i=1,2,3 \cr
T_0(s) &= \int \psi^\dagger {\partial \over \partial s} \psi\ d^4x.}$$
The dimension of $T_i,T_0$ is equal to the number of zero modes $\psi(x,s)$ and
therefore may also jump at $e^{i s}=W_p$.
Away from the jumping points, $T_i(s),T_0(s)$ satisfy Nahm equations \rnahm. The
behavior near the jumping points is generally complicated
\rhurt,
but simplifies when the dimension of $T_i,T_0$ is independent
of $s$
and equal to $N$, the second Chern class of the gauge connection.
This happens when the original gauge connection has no magnetic charges, and
corresponds to the brane configuration considered in section 2.2. In this
special case,
the matrices $T_i,T_0$ have finite left and right limits near $s=s_p\equiv
-i\log W_p,
p=1,\ldots,k$ and satisfy
$$T_0(s_p+0)=T_0(s_p-0),\quad T_i(s_p+0)-T_i(s_p-0)=\rho_i(Q_p).$$
Here $Q_p,p=1,\ldots,k,$ are elements of $\IH^N$, and $\rho(Q)$ is
the moment map for the natural action of $U(N)$ on $\IH^N$ given in eq. \unmap.
These conditions can be combined with the Nahm equations into,
\eqn\nahmwithdelta{{dT_i\over ds}+[T_0,T_i]+{1\over 2}\e_{ijk}[T_j,T_k]=\sum_p
\rho_i(Q_p)\delta(s-s_p).}
The latter equation has the form eq. \momentmap\ and coincides with the
condition of $\D$-flatness in section 2.2, as claimed.

Conversely, given a solution of Nahm equations one can reconstruct the gauge
connection $A$. We are not going to describe the inverse Nahm transform here,
and simply remark that it can be obtained by studying the dynamics of a probe
brane along
the lines discussed in \refs{\rmdf,\rdiaconescu}.

We would like to stress that the correct jumping conditions are absolutely
crucial in order to get the right moduli space. In our discussion
they came out naturally from the analysis of the strings
stretched between D0 and D4-branes. An attentive reader has probably noticed a
close connection between the brane configuration of section 2.2 and that
considered in \rdiaconescu. We want to point out an important difference
between the two.
When we perform a T-duality on the D4-D0 configuration of section 2.2
along the compact direction, we obtain a system of $N$ parallel D1-branes
compactified on a circle with impurities representing the $k$ D3-branes.
On the other hand, the system studied in \rdiaconescu\ contained $N$ D1-branes
suspended between two parallel D3-branes. The suspended
D1-brane corresponds to a monopole in the world-volume theory of the D3-branes.
The moduli space of monopoles is described by Nahm equations on an interval
$s\in (0,1)$ where the $T_i$ matrices have poles at the ends of the interval
\refs{\rnahm,\rhitchin}.
It was shown in \rdiaconescu\ that Nahm equations indeed follow naturally
from an analysis of the dynamics of D1-branes, with $s$ parallel to the 
world-volume
of the D1-branes. However in the case of suspended $D1$-branes,
it seems  difficult
to derive the right boundary conditions for the $T_i$ matrices from
considerations of brane
dynamics. In principle, the correct boundary conditions should be derivable
from
our
setup by taking a limit where some of suspended D1-brane segments are taken
to infinity in the $x^2,x^3,x^4$ directions.

We now turn to the less familiar and more interesting case of $U(k)$ instantons
on $\IR^2\times T^2$.
By an instanton, we mean a connection $A$ with self-dual curvature $F_A$
and finite action. In particular, this means that the second Chern class is
well-defined.
We allow for non-trivial Wilson lines at infinity breaking $U(k)$ down to
$U(1)^k$. The possible Wilson lines are parametrized by $k$ unordered points
$z_1,\ldots,z_k$ on the dual torus $\wT^2$. No magnetic
charges are allowed because with only two noncompact dimensions, the magnetic
field of
a monopole would decay as $1/r$ and the action would diverge logarithmically.
Let us denote
the holomorphic coordinate on $\IR^2$ by $t$ and the holomorphic coordinate
on $T^2$ by $w$. To perform the Nahm
transform,  we again look for solutions of the Dirac equation with auxiliary
$U(1)$ Wilson lines. These auxiliary parameters live on the dual torus
$\wT^2$ whose
complex coordinate we call $z$. For generic values of $z$, the Dirac
equation will have $N$ normalizable zero modes (we assume here that
the connection $A$ on $\IR^2\times T^2$ can be extended to a connection
on the compactified space $S^2\times T^2$). In fact, we expect these zero
modes to decay exponentially at infinity. Then we can define Nahm matrices
living on $\wT^2$,
$$ A_z=\int \psi^\dagger {\partial\over\partial z}\psi\, d^2t\,d^2w,\quad
\Phi=i\int \psi^\dagger\, t\, \psi\, d^2t\,d^2w.$$
Using arguments along the lines of \refs{\rnahm,\rcg}, it can be shown that 
$A_z$ and $\Phi$
satisfy Hitchin equations away from $z=z_p,p=1,\ldots,k.$ As $z\ra z_p$ the
norm of
some zero modes of the Dirac operator diverges, and therefore  $A_z$ and
$\Phi$
diverge as well. We therefore expect that $A_z$ and $\Phi$ will have poles at
$z=z_p,p=1,\ldots,k$.

The argument of the preceeding paragraph shows that the Nahm transform of a
self-dual $U(k)$
connection $A$ on
$\IR^2\times T^2$ with second Chern class $N$ is a solution of the $U(N)$
Hitchin
equations on the dual torus $\wT^2$ with poles at
$z=z_p,p=1,\ldots,k$. The
position of the punctures encodes the asymptotic eigenvalues of the $U(k)$
Wilson lines.
This is in complete agreement with eq. \hitchin\ obtained from studying brane
dynamics.
We have not established the precise kind of the singularities that $A_z$ and
$\Phi$ should
have, but we believe that we have presented enough evidence that eq. \hitchin\
correctly
describes the moduli space of instantons on $\IR^2\times T^2$. Granted this,
eq.
\hitchin\ tells us what kind of singularities the Nahm matrices must have. For
example, $\Phi$ has a pole at $z=z_p$ with residue proportional to $Q^p\otimes
\tilde{Q}^p$.
It remains an interesting challenge to derive this result using either the Nahm
transform \rnahms\ or twistor methods \rhitchin.

\subsec{Some examples of moduli spaces}

Let us discuss some simple examples of instanton moduli spaces on $\IR^3\times
S^1$
and $\IR^2\times T^2$. We remind the reader that we are only
discussing
instanton configurations with no magnetic charges.

First we give examples of moduli spaces of instantons on $\IR^3\times S^1$.
Some interesting recent discussion of these instanton solutions appeared in
\rkraan. For $N=1$, the Nahm equations require that $T_i$ be locally constant
with
discontinuities
occuring at $s=s_p,p=1,\ldots,k.$ The center of mass degrees of freedom
clearly live
in $\IR^3\times S^1$, while the rest of the moduli space can be shown to be a
$4k-4$-dimensional manifold
endowed with a `periodic' version of the Lee-Weinberg-Yi metric
\refs{\rweinberg,\rleeyi}. It has $k-1$ compact
directions and a singularity at the origin. In particular, for
$k=2$ it
reduces to a ${\IZ_2}$ orbifold of the Taub-NUT metric. It is also interesting
to look
at the limit when $k'\leq k$ of the points $z_p$
coincide. This is also the limit in which the broken gauge group $U(k)$ is
partially restored
to $U(k')$. In this limit, $k'-1$ compact directions of the moduli space
decompactify but the moduli space description seems to remain valid \rweinberg.
Note that turning on the FI deformation will smooth out the singularity of the
moduli space.

For $N=2$, the Nahm equations can be solved in terms of elliptic functions; see
\rdancer\ for example.
The resulting moduli space splits into a product of $\IR^3\times S^1$ and
a fairly complicated $8k-4$-dimensional hyper\kh manifold. We have not
investigated the detailed properties of this manifold; we simply remark that
it has $2k-1$ compact directions. In this case, it seems that the limit where
$U(k')$ symmetry is restored does not lead to decompactification of the moduli
space describing the relative motion of the D1-branes. This is in accord with
our intuition from the matrix model interpretation, which we will discuss
in section four. In general, it is easy to see that the moduli space of
$N$ $U(k)$ calorons with no magnetic charges has dimension $4Nk$ and a total 
of $Nk$ compact directions.

Now let us turn to instantons on $\IR^2\times T^2$. Here we only discuss the
case $N=1$. The situation is quite different from the case of instantons on 
$\IR^3 \times S^1$. The solutions of the Hitchin equations with impurities
depend on $4k$ parameters. However, not all of the parameters correspond to
normalizable zero modes. The mathematical reason goes as follows: the Hitchin 
equations are abelian in this case and easily solved in terms of elliptic
functions with simple poles at $z=z_p$ for $p=1,\ldots, k$. The residues of 
these elliptic functions are determined by the 
variables $Q$ and $ \tilde{Q}$. Therefore, the tangent vectors associated to 
varying $Q$ and $ \tilde{Q}$ also have simple poles. The norm of these tangent
vectors is then logarithmically divergent. In this case where $N=1$, the only
finite norm tangent vector corresponds to a constant solution. It is then
easy to see that the moduli space is just $\IR^2 \times T^2$. 

In the limit where two or more impurities collide,
so some non-abelian gauge symmetry is restored, the dimension of the moduli
space becomes larger. This can be seen by noting that even if we require the
residue at the collision point to remain fixed, $Q$ and $\tilde{Q}$ have a 
non-trivial moduli space when $k>1$. This moduli space is the same as the 
Higgs branch of a $U(1)$ gauge theory with $k$ electrons.  

For $N>1$, the story is similar: $4k-4$ zero modes freeze out and the 
dimension of the resulting moduli space is $4(Nk-k+1)$ \refs{\rnew, \rkap}. 
When some impurities
collide, the dimension of the moduli space will again jump. It would be 
very interesting to analyze the structure of the moduli space for $N>1$
in detail.

\newsec{Some Applications of Impurity Models}

\subsec{Matrix models for $SU(k)$ Yang-Mills theories}

A natural way to construct N=4 $U(k)$ Yang-Mills is to compactify $k$
parallel M theory five-branes on $T^2$. The $U(1)$ describing the center of mass
motion effectively decouples, leaving an $SU(k)$ theory. At energies well below the
eleven-dimensional Planck mass $M_{pl}$ and the scale set by the torus,
the theory reduces to Yang-Mills in four dimensions. The complex structure of
the torus determines the coupling of the Yang-Mills theory, and S-duality is
therefore made manifest \rwtensor. In this section,
we will describe the DLCQ description of $k$ parallel M theory five-branes
wrapped on either $S^1$ or $T^2$. The DLCQ description involves the Higgs
branches of the impurity models that we have considered \rnew.

Longitudinal five-branes are described in matrix theory by D4-branes \rbd. To
obtain a matrix formulation of $k$ compactified five-branes in the sector with
$P^+ = N/\RV$, we need to
consider the dynamics of $N$ D0-branes in the presence of $k$ wrapped
D4-branes.
We will follow the usual prescription \rlimit. Longitudinal five-branes
wrapped on $S^1$ or $T^2$ give longitudinal D4 and D3-branes, respectively.
Therefore, longitudinal D4 or D3-branes are represented by impurities in the
$1+1$ and $2+1$-dimensional field theories describing D0-branes on $S^1$ and
$T^2$. The spacetime transverse to the longitudinal branes is described by 
the Coulomb branch of the impurity
theories, where $Y$ has non-zero expectation value. The compactified $(2,0)$
theory in DLCQ
is then described by the other branch of the impurity theory -- the Higgs branch. 
This is analogous
to the uncompactified $(2,0)$ case \refs{\rquantumfive, \rstringfive}.
Note that if we turn on the FI deformation, the Coulomb branch is lifted
and so the Higgs branch decouples from spacetime. 

Let us consider the more interesting case of  $SU(k)$ N=4 Yang-Mills on
parallel
longitudinal D3-branes. Let $R_1, R_2$ be the radii of the torus $T^2$ on which
we
compactify the five-branes. For simplicity, let us assume the torus is
rectangular.
At energies $E << 1/R_1, 1/R_2$, the effective theory on the
compactified five-branes is Yang-Mills. We hold fixed the Yang-Mills coupling,
\eqn\ymcoupling{ g^2_{YM} = {R_1\over R_2},}
which is the same as the type IIB string coupling. When the area of the torus
is finite,
we are describing the type IIB string in DLCQ
with transverse space $\IR^7 \times S^1$, where the circle has 
size $1\over M_s^2 R_2$. The longitudinal
D3-branes are transverse to the compact circle. The string scale
is given in terms of the eleven-dimensional Planck scale: $ M_s^2 = M_{pl}^3
R_1$.

The matrix model for this system is the $2+1$-dimensional impurity model
of section 2.3 with coupling constant,
\eqn\matrix{ g^2_{mat} =  {\RV \over R_1 R_2},}
The theory lives on the dual torus with sides $\S_1, \S_2$ where,
$$ \S_i = {1\over M_{pl}^3 \RV R_i}.$$
Since we are considering energies far below $M_s$, we can ignore modes with
a non-trivial dependence on the two spatial directions of this dual torus. Our
$2+1$-dimensional
theory then reduces to quantum mechanics on the Higgs branch moduli space.
Note that S-duality is
manifest in this formulation. The coupling constant in the quantum
mechanical sigma model diverges,
$$ g^2_{QM}  = {g^2_{mat}\over \S_1 \S_2 }\r \infty,$$
so we are left with a conformal theory on the Higgs branch. There
are
compact directions on the moduli space, but the energies associated to
excitations
along those directions are much higher than the energies we are considering
\rnew.
We can therefore take wavefunctions which are slowly varying along the
compact
directions. This is the matrix model analogue of ignoring Kaluza-Klein
modes along $T^2$.

The case when the impurities are separated on the dual torus corresponds to
Yang-Mills with light-like Wilson lines. These light-like Wilson lines are
determined by points on $T^2$ rather than $S^1$ because
four-dimensional Yang-Mills in light-cone is effectively three-dimensional.
Therefore the compact
scalars dual to the three-dimensional gauge fields can also be given
expectation values. Note that as
$N\r \infty$, the light-like Wilson lines should disappear and we should
recover a description of the interacting conformal field theory.

 We noted in the previous section that certain zero modes 
present in the 
$S^1$ case are frozen out in the $T^2$ case.  Recall that some zero modes seem 
to be restored when two or more impurities collide. This
corresponds to the limit where some non-abelian gauge symmetry is restored. We
should then optimistically be describing the interacting conformal
field theory on coincident D3-branes in DLCQ at finite $N$. If we turn on a 
FI deformation, the 
moduli space is smooth, even in the case of coincident impurities. It is
important to determine whether this matrix description
actually captures all the degrees of freedom needed to describe the
conformal field theory at finite $N$. This depends largely on how much can
be understood about the moduli space of instantons on $\IR^2\times T^2$
with no Wilson lines, or perhaps about its non-commutative generalization.
Similarly, the matrix description of M theory five-branes wrapped on $S^1$
is given by the impurity theory considered in section 2.2.

\subsec{Non-commutative instantons and Coulomb branches}

Impurity theories can be used to solve for the Coulomb branches of certain
conventional N=2 d=4 gauge theories. For the specific model studied in 
section 2.3, the Higgs
branch is related to the Coulomb branch of the four-dimensional theory
obtained by placing $N$ D3-branes on an $A_{k-1}$ singularity \rnew.
This quiver theory has gauge group $U(N)^k$ with a chain of
bifundamentals \doumoo. All the $U(1)$ factors except for the diagonal
$U(1)$ factor will become free in the infra-red. This is the Coulomb branch
analogue of the freezing of moduli noted in section 3.3. There is 
a T-dual
realization of this configuration in terms of $N$ D4-branes wrapped on
a compact circle with $k$ NS five-branes located at points on the
circle \rfour. This configuration is T-dual to an $A_{k-1}$ singularity
when the NS five-branes are coincident.

We have the freedom of
turning  on a FI parameter in our impurity model, which lifts the Coulomb
branch and deforms the Higgs branch. The Higgs branch can then be interpreted
as describing instantons on a non-commutative $\IR^2\times T^2$.
How does this deformation
appear in the quiver theory and its T-dual realization? The quiver theory has
a single mass parameter. Turning on the mass lifts the Higgs branch, which
in the matrix model interpretation describes motion in the transverse spacetime. 
This deformation corresponds
to turning on the FI parameter in the impurity model. In the T-dual
description with D4-branes and NS five-branes, it corresponds to placing the
branes in a non-trivial background \rfour.

Let us discuss in some detail the case of N=2 $U(N)$ gauge theory with a massive
adjoint hypermultiplet. In discussing the relevant brane configurations, we will
follow the conventions of \rfour. The brane configuration contains  
$N$ D4-branes wrapped on a circle with a
single NS five-brane located at a point on the circle parametrized by $x^6$.
The motion of the D4-branes in the directions parallel to the NS five-brane
is parametrized by a complex coordinate $v=x^4+ix^5$. In the case without
the mass deformation, we consider M theory on $T^2$ with some five-branes.
The circle on which we reduce M theory to type IIA is parametrized by $x^{10}$.
To turn on the mass deformation, we take an affine $\, \IC$-bundle over the
torus parametrized by $x^6$ and $x^{10}$. The fiber $\,\IC$ is coordinatized
by $v$. As we go around the circle parametrized by $x^6$, $v$ is shifted 
by a complex number $m$. In this case, the low-energy theory on the D4-branes
is $U(N)$ Yang-Mills with an adjoint hypermultiplet with mass $m$. The Coulomb 
branch was determined in \refs{\rdonagi, \rfour}. Using the Higgs branch of 
impurity theories, we can determine
this Coulomb branch in a quite different way.

We consider the theory of section 2.2 with a single D4 wrapped on $T^2$ and $N$
D0-branes. If we do not turn on the FI deformation, the Higgs branch is quite
trivial. It is $S^N(\IR^2 \times T^2)$, and this agrees with the Seiberg-Witten
fibration which describes the Coulomb branch of N=4 $U(N)$ Yang-Mills. The complex
structure of the $T^2$ is determined by the coupling of the Yang-Mills theory. If
we turn on the FI deformation, the Higgs branch becomes non-trivial and is given
by the moduli space of solutions of eq. \hitchin\ with $k=1$, 
\eqn\nodon{ \eqalign{ & F_{z\oz}-[\Phi,\Phi^\dagger]={1\over
2R_1R_2}
\delta^2(z)\left( Q\otimes Q^{\dagger } -\tilde{Q}^{\dagger }
\otimes\tilde{Q}\right),\cr}}
\eqn\don{ \eqalign{ 
& \overline{D}\Phi=-{1\over 2R_1R_2} \delta^2(z) \left(
Q\otimes\tilde{Q} - \xi \right). \cr}}
The FI deformation $\xi$ of the impurity theory is proportional to the mass
deformation $m$ of the Coulomb branch. To compare with the solution of \rdonagi, 
let us consider eq. \don\ modulo the action of the complexified gauge group. This
is sufficient to determine the complex structure of the Seiberg-Witten fibration.
First consider the trace of eq. \don. This gives, 
\eqn\trace{ \eqalign{ 
& \overline{\partial} \, \Tr \Phi=-{1\over 2R_1R_2} \delta^2(z) \left(
\tilde{Q}Q - N \xi \right). \cr}}
This implies that $\Tr \Phi$ is a meromorphic function on $T^2$ with a single
pole at $z=0$. The residue must therefore be zero, forcing:
\eqn\residue{ \tilde{Q}Q = N \xi.}
The system described by eq. \don\ and eq. \residue\ has been considered in 
\rnikt, 
and shown to be equivalent to the (complexified) elliptic Calogero-Moser system.
In turn, the  Calogero-Moser system is equivalent to the Donagi-Witten solution 
\rmor. Concretely, this means the following: the Calogero-Moser system is 
integrable, so it has $N$ Poisson commuting integrals of motion. There are 
$N$ pairs of action-angle variables, and the phase space is therefore fibered by
tori of angle variables. After complexification, this fibration becomes equivalent
to the Seiberg-Witten fibration describing softly broken N=4 $U(N)$ Yang-Mills. 
This approach to solving N=2 d=4 gauge theories will be further explored in \rkap.

\bigbreak\bigskip\bigskip\centerline{{\bf Acknowledgements}}\nobreak

It is our pleasure to thank S. Cherkis, O. Ganor, G. Moore,
N. Nekrasov, J. Park, A. Uranga, C. Vafa
and E. Witten for helpful discussions. The work
of A.K. is supported by DOE DE-FG02-90ER40542; that of S.S. by NSF
grant DMS--9627351.

\listrefs

\end